\documentstyle[preprint,aps]{revtex}

\tightenlines

\begin{document}

\draft

\title{Meta-Plaquette Expansion for\\
the Triplet Excitation Spectrum in CaV$_{\bbox{4}}$O$_{\bbox{9}}$}

\author{Zheng Weihong\cite{byline1}, J. Oitmaa\cite{byline2}, and
 C. J. Hamer\cite{byline3} }
\address{
School of Physics,                                              
The University of New South Wales,                                   
Sydney, NSW 2052, Australia}                      

\date{July 13, 1998}


\maketitle 

\begin{abstract}
We study antiferromagnetic, $S=1/2$ Heisenberg models with 
nearest and second neighbor interactions
on the one-fifth depleted square lattice which describes
the spin degrees of freedom in the spin-gap system CaV$_4$O$_9$.
The meta-plaquette expansion for the triplet excitation spectrum is
extended to fifth order, and the results  are compared with experimental data on
CaV$_4$O$_9$.
We attempt to locate the phase boundary between magnetically ordered and
gapped phases.
\end{abstract}                                                        

\pacs{PACS numbers: 75.10.-b., 75.10.Jm, 75.40.Gb, \\ \\  \\
Submitted to  Phys. Rev. B}

\narrowtext


The physics of low-dimensional Heisenberg antiferromagnets has been
a subject of much interest. These models represent a class of
quantum many-body Hamiltonians which accurately model real materials
and whose properties can be calculated by a number of controlled
analytical and numerical methods. 

Recently, the observation of spin gaps in the quasi-2D material
CaV$_4$O$_9$\cite{taniguchi,oha97,kod96,kod97} has attracted much attention 
from theorists.
In this layered material, $S=1/2$ vanadium atoms form a 
one-fifth depleted square lattice which we will
refer to as the CAVO lattice, and which is shown
in Fig.~\ref{CAVOlattice}.
It is expected that low energy properties of CaV$_4$O$_9$ 
are described by the antiferromagnetic Heisenberg
model with four kinds of exchange interactions:
$J_1$ and $J'_1$ for the nearest-neighbor intra- and 
inter-plaquette coupling respectively,
and  $J_2$ and $J'_2$ for the second-neighbor interaction,
as shown in Fig. 1.
One is thus led to consider the following Hamiltonian:
\begin{eqnarray}
H= && J_1\sum_{(i,j)} {\bf S}_i \cdot  {\bf S}_j+
J'_1\sum_{(i,k)} {\bf S}_i \cdot  {\bf S}_k \nonumber\\
&&+J_2\sum_{(i,l)} {\bf S}_i  \cdot {\bf S}_l +
J'_2\sum_{(i,m)} {\bf S}_i  \cdot {\bf S}_m,
\label{Ham}
\end{eqnarray}
where the sums run over nearest-neighbor bonds within 
plaquettes ($J_1$),
nearest-neighbor bonds between plaquettes ($J'_1$), 
second-neighbor bonds within plaquettes ($J_2$),
and second-neighbor bonds between plaquettes ($J'_2$).
This Hamiltonian has the property that every
spin is equivalent, just as every vanadium atom in CaV$_4$O$_9$
is equivalent. 

To understand the mechanism of the gap formation,
a variety of methods, including quantum Monte Carlo simulation\cite{troyer,katoh},
perturbative and high-temperature series 
expansion\cite{kod97,zwh_ser,gel97,tak96,sano,fuk98}, 
exact diagonalization\cite{sano,alb96}, DMRG method\cite{whi96},
spin-wave theory\cite{fukumoto}, and other
techniques\cite{sachdev,miy96,kat97} have been applied to this model. Based on simple
considerations, most authors have assumed $J_1\simeq J'_1$,
$J_2\simeq J'_2$, and $J_1 > J_2$, and with this assumption,
the experimental data is best supported by assuming that the second-neighbor
interactions are roughly half the first-neighbor interactions:
but the fits are not entirely satisfactory.  

Recently,  Pickett\cite{pic97} has used LDA calculations to
study the magnetic properties of CaV$_4$O$_9$.
He argues that out-of-plane distortions in
the arrangement of the vanadium atoms are quite important. 
The one-fifth depleted square lattice of vanadium atoms is actually 
made up of two (planar) layers, slightly above and below one 
$ab$-plane, and each forming
a square lattice of ``meta-plaquettes.''
These meta-plaquettes are 
to be distinguished from the plaquettes considered before 
by Ueda {\it et al.}\cite{ueda} and other authors in 
the ``plaquette RVB'' (PRVB) scenario. 
The plaquettes involve nearest-neighbor spins, two
of them lying in the upper layer and two in the lower layer. In this
case, a significant second-neighbor interaction is expected within
the plaquettes. In contrast, the meta-plaquettes of Pickett lie
either entirely in the upper or entirely in the lower layer. 
Pickett argues that although the spins
within a meta-plaquette are further apart than spins between two neighboring
meta-plaquettes, various quantum-chemical arguments conspire to make
the interactions within meta-plaquettes the strongest.
He finds that to leading approximation
the magnetic system should be regarded as decoupled meta-plaquettes.
One important point to note is that within a meta-plaquette the
interactions are expected to be primarily between neighboring spins only.

On the experimental side,
a neutron inelastic scattering experiment has been recently carried
out by Kodama {\it et al}\cite{kod97}, and the triplet excitation spectrum
was obtained. They also reported that the theoretical
results using the second order perturbation from
the disconnected meta-plaquette model, which is defined
by $J_1=J'_1=J_2=0$, agrees with the experimental results
when $J'_2=14.73$, $J_1=J'_1=5.76$, and $J_2=1.25$(meV).
(In their fitting procedure,
$J_1=J'_1$ is assumed.). However, there are some mistakes in their
calculation. These mistakes have been corrected by 
Fukumoto and Oguchi\cite{fuk98}, and they also extend the results to
third order for given values of $J'_1/J_1$ and $J_2/J_1$.
In the third order perturbation expansion,
the exchange parameters for CaV$_4$O$_9$ are determined as
$J'_2=14.0(1)$, $J_1=J'_1=6.8(2)$, and $J_2=1.7(2)$(meV).
Using these exchange parameters and the technique of numerical
diagonalization  of finite lattices,
Takano and Sano\cite{tak98} calculated the magnetic susceptibility $\chi$, 
and found the experimental data to be about 30\% smaller than the
theoretical results. They interpret this as due to the  30\% volume 
fraction of nonmagnetic components in the CaV$_4$O$_9$ sample, and  this 
explanation is also consistent with other experimental data.
In this report, we extend the meta-plaquette series expansions 
for the triplet excitation spectrum to fifth order.

To carry out the expansion about disconnected meta-plaquettes,
we rewrite the Hamiltonian in Eq. (\ref{Ham}) as:
\begin{equation}
H/J'_2 = H_0 + x V
\end{equation}
where $H_0$ and $V$ are the unperturbed Hamiltonian and perturbation, respectively, and
$x$ is the expansion parameter. They are defined as:
\begin{eqnarray}
H_0 &=&  \sum_{(i,m)} {\bf S}_i  \cdot {\bf S}_m \nonumber \\
V &=& \sum_{(i,j)} {\bf S}_i \cdot  {\bf S}_j+
y_1 \sum_{(i,k)} {\bf S}_i \cdot  {\bf S}_k 
+ y_2 \sum_{(i,l)} {\bf S}_i  \cdot {\bf S}_l \nonumber \\
x &=& J_1/J'_2   \nonumber \\
y_1 &=& J'_1/J_1 \nonumber \\
y_2 &=& J_2/J_1
\end{eqnarray}

We use the linked-cluster expansion method which has been previously reviewed
in several articles\cite{he90,gel90,gelmk}.
Here we have performed a three parameter ($x$, $y_1$ and $y_2$) 
meta-plaquette
series expansion
for the triplet elementary excitation spectrum up to fifth order.
The excitation spectrum takes the form:
\begin{equation}
\Delta (k_x, k_y)/J'_2 = \sum_{i=0}^5 \sum_{j,k,m,n} a_{i,j,k,m,n} x^i y_1^j y_2^k
 \left[ \cos(m k_x + n k_y) + \cos( n k_x - m k_y) \right] /2
\end{equation}
Where $a_{i,j,k,m,n}$ are the expansion coefficients.
Up to order $i=5$, there are  a total of 698 non-zero terms in 
the series for general $(k_x,k_y)$:
since this requires an inordinate amount
of space to reproduce in print, it is available from the authors
on request. 
Instead, in Table I, we list the series for fixed  $(k_x,k_y)=(0,0)$ and $(\pi,\pi)$.
Our results agree with the results of  
Fukumoto and Oguchi\cite{fuk98} and extend the series by two orders.
We also agree with  the results of 
Gelfand and Singh\cite{gel97} who only considered the $(0,0)$ gap in
the case  $J_1=J'_1$ and $J_2=0$.


We first seek to refine the values of the exchange parameters for
CaV$_4$O$_9$ using our extended series. Fig. 2 shows the dispersion
curve, based on the parameter values 
$J'_2=14.0$, $J_1=J'_1=6.8$ and $J_2=1.7$(meV)
of Fukumoto and Oguchi\cite{fuk98}, evaluated by direct summation of
the expansion to 2nd, 3rd, 4th and 5th order. The experimental
points of Kodama {\it et al.}\cite{kod97} are also shown.
It is clear from the figure that the higher terms are small, for 
this region of parameters, and the curves are almost indistinguishable,
except near the point $(\pi,\pi)$. Further experimental data at
this point would allow a refinement of the parameters,
but without this there is little basis for doing so.

Within the $(x, y_1, y_2)$ parameter space there will be a critical
surface separating a phase with long range antiferromagnetic
order (at $T=0$) from a gapped spin-liquid phase (in which
the real material CaV$_4$O$_9$ lies). Although a 6-term
series is really too short to determine critical points with any
accuracy, we are able to make a first estimate of this, at least
in the region where the minimum gap occurs at $(0,0)$.
To do this we fix values of $y_1$ and $y_2$ and analyze
Dlog Pad\'{e} approximants\cite{gut} to the series for
$\Delta (x)$ which vanishes at $x_c$. Figure 3 shows slices
of the critical surface in the $(x, y_2)$ plane for various choices
of $y_1$. The lines separate an upper gapless antiferromagnetic
phase from the lower gapped phase. The real material has 
$x=0.49$, $y_1=1$, $y_2=0.25$ and lies well within the gapped phase.
The vanishing of the gap is characterized by an exponent $\nu$,
which we estimate as $\simeq 0.7$. This suggests that the 
transition may lie in the universality class of the classical
$d=3$ Heisenberg model.

Finally we investigate the qualitative form of the dispersion
curve for a range of parameter values within the gapped phase.
Fig. 4 shows this for fixed $x=0.4$, $y_1=1$ and various
$y_2$. For small $y_2$ the shape of the dispersion curve is
broadly similar to the experimental one, with a minimum at
$(0,0)$ and a maximum at $(\pi,\pi)$. For $y_2\simeq 0.5$
the dispersion along the $k_x$ and $k_y$ directions becomes
quite flat and for $y_2>0.5$ the minimum shifts from
$(0,0)$ to $(\pi,0)$. It is interesting to speculate on whether
these regions could be achievable in CaV$_4$O$_9$
by suitable modification of the structure or by
external strains.

In conclusion we have extended the meta-plaquette
expansion for the triplet excitation spectrum of the Heisenberg 
model on the CAVO lattice by two terms, to 5th order.
We are able to locate at least part of the critical surface
separating gapless and gapped phases. The shape of the 
dispersion curve shows interesting variation with $J_2$, the
intra-plaquette second-neighbor exchange parameter, for
fixed values of the other parameters. In the region of
parameter space for the real material CaV$_4$O$_9$
the contribution of the higher terms is quite small and we are
unable to further refine the values of proposed by
Fukumoto and Oguchi. Further experimental data particularly
around the $(\pi,\pi)$ point would allow this to be done.

This work forms part of a research project supported by a grant
from the Australian Research Council. The computation has been performed
on Silicon Graphics Power Challenge and Convex machines. We thank the New
South Wales Centre for Parallel Computing for facilities and assistance
with the calculations.


\begin{figure} 
\caption{The CAVO lattice, with sites indicated by circles.
The full and open circles represent vanadium ions slightly above and below
the plane, respectively.
The couplings $J_1$, $J'_1$, $J_2$, and $J'_2$ are indicated
by thick solid, thin solid, thin dashed, and thick dashed
lines, respectively.  Note that the meta-plaquette centers lie
on a square lattice with spacing $b$ which is ${\protect\sqrt 5}$
times  the distance between nearest-neighbor sites.  
In characterizing the excitation spectrum
we take $b=1$ and rotate the coordinate system
so that the lines between nearest-neighbor meta-plaquette centers
define the $x$ and $y$ axes.}
\label{CAVOlattice}
\end{figure}

\begin{figure} 
\caption{Triplet excitation spectra along high-symmetry directions
as estimated by direct sum of meta-plaquette expansions to 
order $x^2$ (dotted line),  $x^3$ (short dashed  line),
 $x^4$ (long dashed  line), and $x^5$ (solid  line), 
for $J_1=J'_1=6.8$, $J_2=1.7$ and $J'_2=14.0$(meV).
Also shown as full points are the experimental data by
Kodama {\it et al.}\protect\cite{kod97}}
\label{fig:dispsum2}
\end{figure}

\begin{figure} 
\caption{Partial phase boundary for $y_1=0,0.5,1$ of
 the CAVO lattice Heisenberg
model, as determined from meta-plaquette expansions.
The points with error bars and a solid line to guide
the eye indicate  the phase boundary where the
$(0,0)$ gap vanishes.
The cross shows the position of CaV$_4$O$_9$
in the parameter space.
}
\label{fig:phasdiag}
\end{figure}

\begin{figure} 
\caption{Triplet excitation spectra along high-symmetry directions
as estimated by Pad\'{e} approximants to the series to 
order $x^5$, 
for $y_2=0$, 0.25, 0.5, and 1 (as shown) at $x=0.4$, $y_1=1$.}
\label{fig:dispsum}
\end{figure}

\widetext

\setdec 0.00000000000
\begin{table}
\squeezetable
\caption{Series coefficients for the meta-plaquette expansion of the triplet
excitation spectrum 
$\Delta (k_x, k_y)/J'_2 =  \sum_{i,j,k} 
a_{i,j,k}(k_x,k_y) x^i y_1^{j} y_2^k $. 
Nonzero coefficients $a_{i,j,k}$ for $(k_x,k_y)=(0,0)$ and $(\pi,\pi)$  
up to order $i=5$ are listed.}\label{tabdimgap}
\begin{tabular}{rrr|rrr}
\multicolumn{1}{c}{$(i,j,k)$} &\multicolumn{1}{c}{$a_{i,j,k}(0,0)$}&\multicolumn{1}{c|}{$a_{i,j,k}(\pi,\pi)$} &
\multicolumn{1}{c}{$(i,j,k)$} &\multicolumn{1}{c}{$a_{i,j,k}(0,0)$}&\multicolumn{1}{c}{$a_{i,j,k}(\pi,\pi)$} \\
\hline 
 ( 0, 0, 0) &\dec   1.000000000  &\dec   1.000000000 &( 4, 1, 3) &\dec $-$1.274215356$\times 10^{-1}$  &\dec   1.274215356$\times 10^{-1}$ \\
 ( 1, 0, 0) &\dec $-$1.333333333  &\dec   1.333333333 &( 4, 2, 0) &\dec   1.074058127$\times 10^{-1}$  &\dec $-$7.455165932$\times 10^{-2}$ \\
 ( 1, 0, 1) &\dec   6.666666667$\times 10^{-1}$  &\dec   6.666666667$\times 10^{-1}$ &( 4, 2, 1) &\dec   2.690073898$\times 10^{-1}$  &\dec $-$1.082138472 \\
 ( 1, 1, 0) &\dec   6.666666667$\times 10^{-1}$  &\dec $-$6.666666667$\times 10^{-1}$ &( 4, 2, 2) &\dec   3.365078472$\times 10^{-2}$  &\dec $-$6.298532394$\times 10^{-1}$ \\
 ( 2, 0, 0) &\dec $-$4.050925926$\times 10^{-1}$  &\dec   3.935185185$\times 10^{-2}$ &( 4, 3, 0) &\dec $-$8.134986977$\times 10^{-2}$  &\dec $-$6.022038137$\times 10^{-3}$ \\
 ( 2, 0, 1) &\dec   8.888888889$\times 10^{-1}$  &\dec $-$8.888888889$\times 10^{-1}$ &( 4, 3, 1) &\dec $-$7.119885055$\times 10^{-2}$  &\dec   2.658296322$\times 10^{-1}$ \\
 ( 2, 0, 2) &\dec   6.597222222$\times 10^{-2}$  &\dec   6.597222222$\times 10^{-2}$ &( 4, 4, 0) &\dec   1.385933255$\times 10^{-2}$  &\dec   3.384182408$\times 10^{-2}$ \\
 ( 2, 1, 0) &\dec   8.564814815$\times 10^{-2}$  &\dec   8.564814815$\times 10^{-2}$ &( 5, 0, 0) &\dec $-$2.353587765$\times 10^{-1}$  &\dec   9.367446624$\times 10^{-2}$ \\
 ( 2, 1, 1) &\dec $-$4.444444444$\times 10^{-1}$  &\dec   4.444444444$\times 10^{-1}$ &( 5, 0, 1) &\dec   4.839207165$\times 10^{-1}$  &\dec   2.629675619$\times 10^{-1}$ \\
 ( 2, 2, 0) &\dec   6.597222222$\times 10^{-2}$  &\dec $-$3.229166667$\times 10^{-1}$ &( 5, 0, 2) &\dec $-$9.949929589$\times 10^{-1}$  &\dec   7.710126890$\times 10^{-1}$ \\
 ( 3, 0, 0) &\dec $-$2.817804784$\times 10^{-1}$  &\dec   1.440972222$\times 10^{-1}$ &( 5, 0, 3) &\dec   9.187244689$\times 10^{-2}$  &\dec $-$3.584889115$\times 10^{-2}$ \\
 ( 3, 0, 1) &\dec   8.875064300$\times 10^{-2}$  &\dec $-$3.012956533$\times 10^{-1}$ &( 5, 0, 4) &\dec $-$8.793820229$\times 10^{-2}$  &\dec   8.793820229$\times 10^{-2}$ \\
 ( 3, 0, 2) &\dec $-$3.766718107$\times 10^{-1}$  &\dec   3.766718107$\times 10^{-1}$ &( 5, 0, 5) &\dec   1.620527344$\times 10^{-2}$  &\dec   1.620527344$\times 10^{-2}$ \\
 ( 3, 0, 3) &\dec   7.607542438$\times 10^{-3}$  &\dec   7.607542438$\times 10^{-3}$ &( 5, 1, 0) &\dec   4.041080312$\times 10^{-1}$  &\dec $-$4.210403484$\times 10^{-1}$ \\
 ( 3, 1, 0) &\dec   3.118730710$\times 10^{-1}$  &\dec $-$2.667582948$\times 10^{-1}$ &( 5, 1, 1) &\dec $-$4.870132049$\times 10^{-1}$  &\dec $-$4.980867393$\times 10^{-1}$ \\
 ( 3, 1, 1) &\dec   5.377121914$\times 10^{-2}$  &\dec   1.100983796$\times 10^{-1}$ &( 5, 1, 2) &\dec   1.203414600  &\dec $-$1.113097779 \\
 ( 3, 1, 2) &\dec   1.883359053$\times 10^{-1}$  &\dec $-$1.883359053$\times 10^{-1}$ &( 5, 1, 3) &\dec $-$5.450494770$\times 10^{-2}$  &\dec $-$6.017493539$\times 10^{-2}$ \\
 ( 3, 2, 0) &\dec $-$8.644788452$\times 10^{-2}$  &\dec $-$8.135207690$\times 10^{-2}$ &( 5, 1, 4) &\dec   4.396910114$\times 10^{-2}$  &\dec $-$4.396910114$\times 10^{-2}$ \\
 ( 3, 2, 1) &\dec $-$1.073173868$\times 10^{-1}$  &\dec $-$3.426568930$\times 10^{-1}$ &( 5, 2, 0) &\dec $-$3.524127097$\times 10^{-1}$  &\dec   6.926810815$\times 10^{-1}$ \\
 ( 3, 3, 0) &\dec   7.607542438$\times 10^{-3}$  &\dec $-$8.186246142$\times 10^{-3}$ &( 5, 2, 1) &\dec   1.178079295$\times 10^{-1}$  &\dec $-$4.130088800$\times 10^{-1}$ \\
 ( 4, 0, 0) &\dec $-$2.039154182$\times 10^{-1}$  &\dec $-$2.576864986$\times 10^{-2}$ &( 5, 2, 2) &\dec $-$5.404083450$\times 10^{-1}$  &\dec $-$1.147271044 \\
 ( 4, 0, 1) &\dec   6.358148317$\times 10^{-1}$  &\dec $-$3.967625126$\times 10^{-1}$ &( 5, 2, 3) &\dec   1.206115343$\times 10^{-3}$  &\dec $-$1.006669452 \\
 ( 4, 0, 2) &\dec $-$1.728981197$\times 10^{-1}$  &\dec   1.055808005$\times 10^{-1}$ &( 5, 3, 0) &\dec   2.350225176$\times 10^{-1}$  &\dec $-$5.747589936$\times 10^{-1}$ \\
 ( 4, 0, 3) &\dec   2.548430713$\times 10^{-1}$  &\dec $-$2.548430713$\times 10^{-1}$ &( 5, 3, 1) &\dec   5.681996667$\times 10^{-2}$  &\dec $-$2.997134488$\times 10^{-2}$ \\
 ( 4, 0, 4) &\dec   1.385933255$\times 10^{-2}$  &\dec   1.385933255$\times 10^{-2}$ &( 5, 3, 2) &\dec   1.056067425$\times 10^{-1}$  &\dec   3.714433690$\times 10^{-1}$ \\
 ( 4, 1, 0) &\dec   9.481864669$\times 10^{-2}$  &\dec $-$3.738896059$\times 10^{-2}$ &( 5, 4, 0) &\dec $-$9.802141001$\times 10^{-2}$  &\dec   2.993556466$\times 10^{-1}$ \\
 ( 4, 1, 1) &\dec $-$6.426060778$\times 10^{-1}$  &\dec   5.469359068$\times 10^{-1}$ &( 5, 4, 1) &\dec $-$3.779095786$\times 10^{-2}$  &\dec   2.653503873$\times 10^{-1}$ \\
 ( 4, 1, 2) &\dec   5.705034556$\times 10^{-2}$  &\dec   2.893338312$\times 10^{-2}$ &( 5, 5, 0) &\dec   1.620527344$\times 10^{-2}$  &\dec $-$4.403051591$\times 10^{-2}$ \\
\end{tabular}
\end{table}

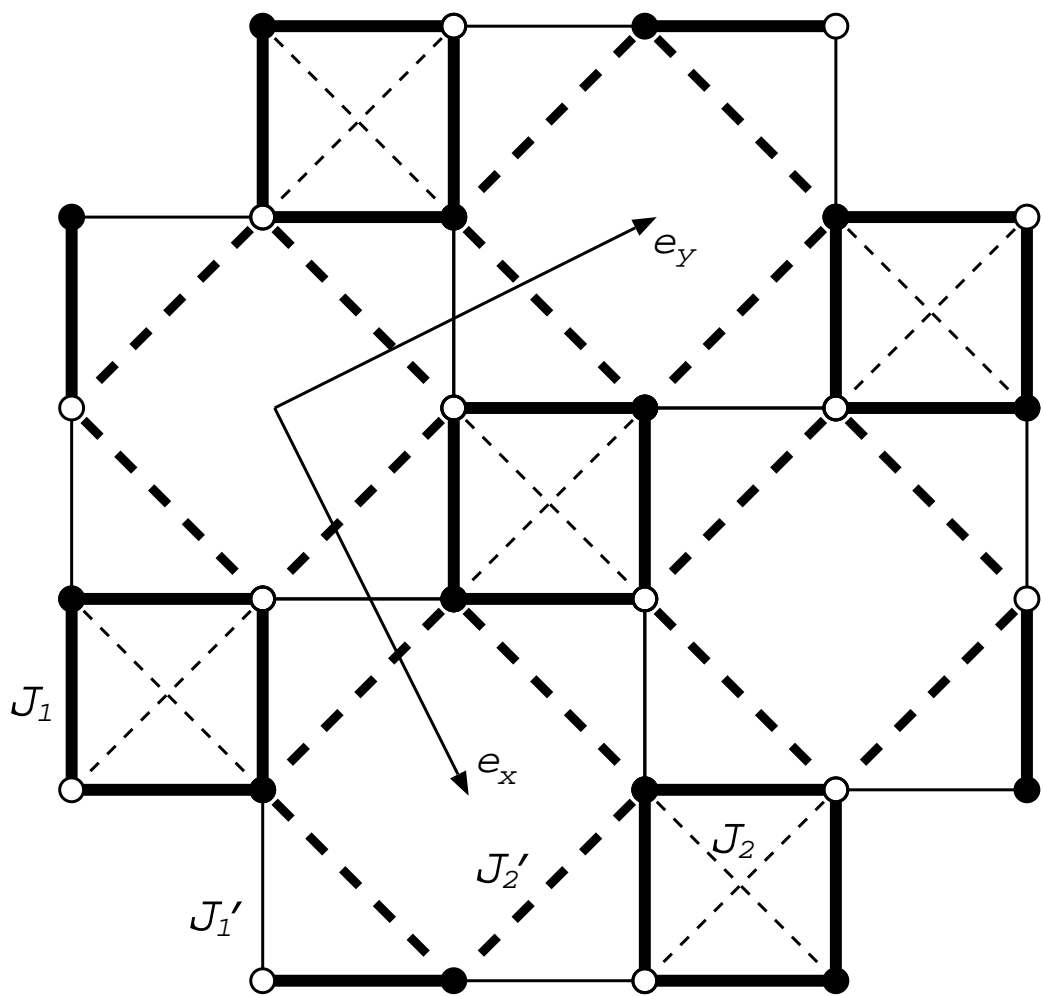
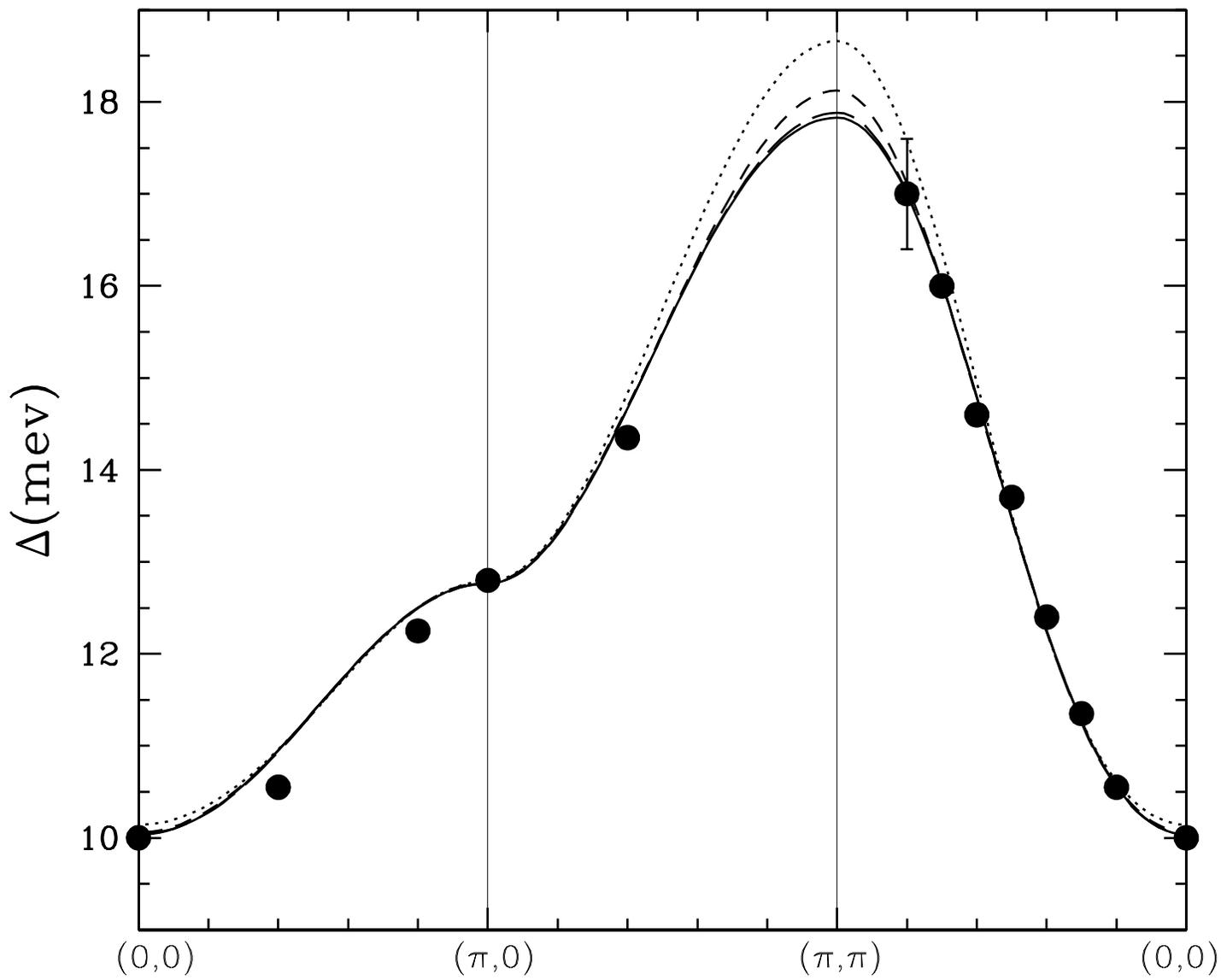
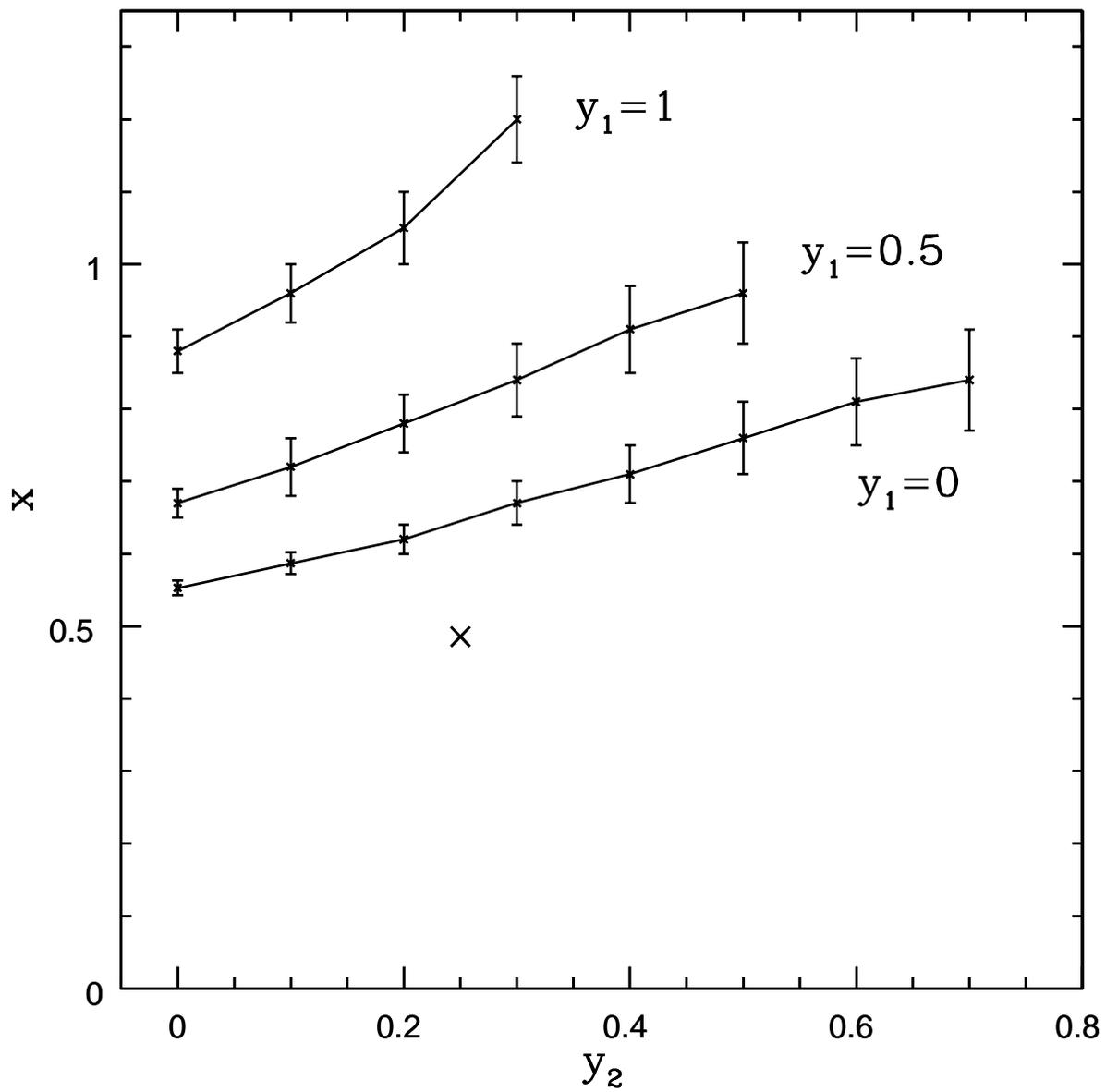
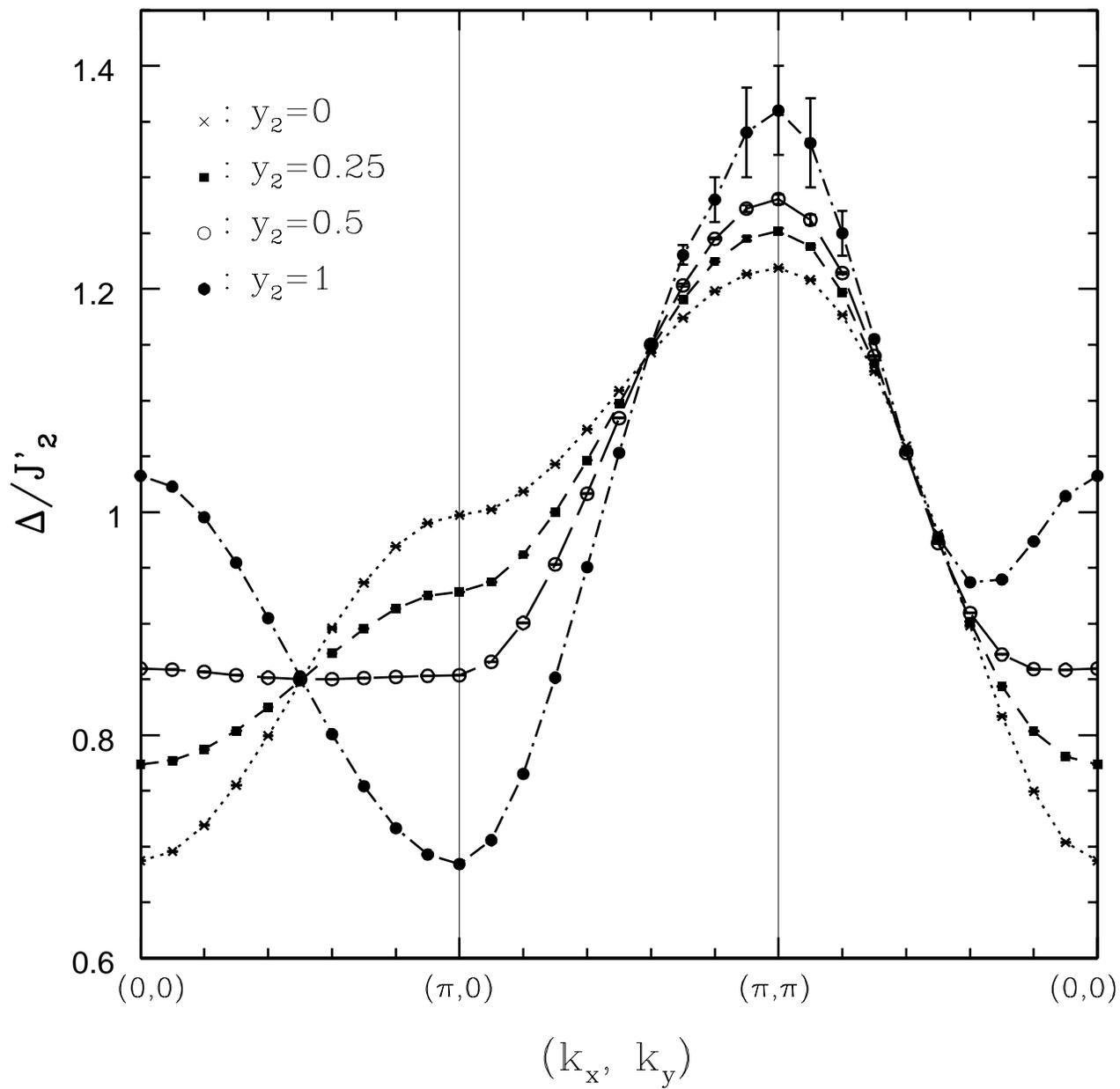
\end{document}